\documentclass[pra, aps, showpacs, superscriptaddress, twocolumn]{revtex4-2}
\usepackage{graphicx,amsmath,amssymb,color}
\usepackage{physics}
\usepackage{graphicx}
\usepackage{epstopdf}
\usepackage{float}
\usepackage{xcolor}
\usepackage{soul}
\usepackage{graphicx}
\usepackage{subcaption}

\usepackage[colorlinks, citecolor={blue!50!black}, urlcolor={blue!50!black}, linkcolor={red!50!black}]{hyperref}

\begin{document}

\title{The Initial Experimental Guide to Implement QKD Polarization Encoding System at Telecom Wavelength}

\author{Zeshan Haider}
 \email{Corresponding author: shani12441@gmail.com}

\date{\today}

\begin{abstract}
The experimental implementation of the polarization encoding system is presented using weak coherent pulses at a low enough frequency. The optical pulses are generated through intensity modulation at the repetition rate of $10$ pulses/sec with the intensity of $7\mu W$, whose polarization is modulated by the Pockel cell. For a $ 5$ km long fiber link, the key rate is thus limited to $ 10$ bits/sec accompanied by temporal synchronization among the Intensity Modulator (IM), Pockel cell, and the gate of the detector. The system was continuously operated for $5$ minutes, and $3000$ binary bits were distributed between two nodes with an error rate less than unity. We have used only a single polarization basis, i.e., horizontal and vertical polarizations, as the objective of this experiment is to validate the feasibility of the setup under simplified conditions, intended primarily for the laboratory-scale demonstrations.

\end{abstract}

\maketitle
\section{introduction}

Quantum Key Distribution (QKD) has emerged as one of the promising techniques for sharing a secret key between two parties, conventionally called Alice and Bob, since its first theoretical protocol was proposed \cite{bennett2014quantum}. Later on, numerous approaches and experiments have been carried out, including QKD through entangled photon pairs, phase-encoded QKD, polarization encoding, continuous variable QKD, QKD in QisKit, among others \cite{ribordy2000long, jennewein2000quantum, enzer2002entangled, ma2007quantum,saeed2022implementation}. The polarization based encoding systems have various advantages over photon-entangled based protocols, as the former systems have high key rates with low decoherence effects. However, polarization drift/error inside the quantum channel induces a large Quantum Bit  Error Rate (QBER) that requires employing efficient error correction codes or real-time compensation of polarization error\cite{chen2009stable,xavier2009experimental,agnesi2019all}. Unlike the free space QKD, it is the birefringence effect in the optical fiber-based QKD systems that causes unpredictable polarization fluctuations. As all QKD protocols give utmost security following the laws of quantum mechanics against the eavesdropper, however, their experimental implementation leaves various loopholes, including imperfect state preparation, channel losses, low detection efficiency limited by dark counts of detectors, among others \cite{brassard2000limitations,zhao2008quantum}. In view of the experiments, single photon states are commonly used through Faint Coherent Pulses (FPC), with the probability of multiphoton states being highly suppressed, though non-zero \cite{kraus2007security,li2014passive,zapatero2025implementation}. The multiphoton states make the QKD system vulnerable to the Photon Number Splitting (PNS) attack. However, if one can compromise this attack, the system can be simplified. The generation and detection of perfectly single photon states are not available to date and demands expensive equipment, leaving the loophole for PNS \cite{ashkenazy2024photon, avanesov2025generalized}. As already stated, this risk of PNS reduces with lowering the mean photon number in faint coherent pulses and dark counts of the detector, though it cannot be completely eleminated due to its probabilistic nature. With the advent of polarization-maintaining and low-loss fibers, the scheme of decoy state QKD has been extensively used on fiber links to mitigate the PNS attack \cite{avanesov2025generalized,ma2005practical,zhao2006experimental,rosenberg2007long}. In this scheme, various intensity levels of incoming optical pulses are achieved through intensity modulators, and usually, the lowest level is used to send the key while all other levels decoy the eavesdropper \cite{schmitt2007experimental,lu2021intensity}. Recently, the trend of extracting  Single photonic states out of multiphotonic states and optical pulses got ignited through various quantum dynamical systems. These systems include Single Photon Raman Interaction (SPRINT)\cite{rosenblum2016extraction,pasharavesh2024sprint}, Single and multiphoton blockade \cite{huang2018nonreciprocal,hamsen2017two,haider2023multiphoton}, quantum dot based single photon source \cite{thapa2025single,thomas2024deterministic}, among others.

\begin{figure}[htb]
	\includegraphics[width=8.5cm,height=4cm]{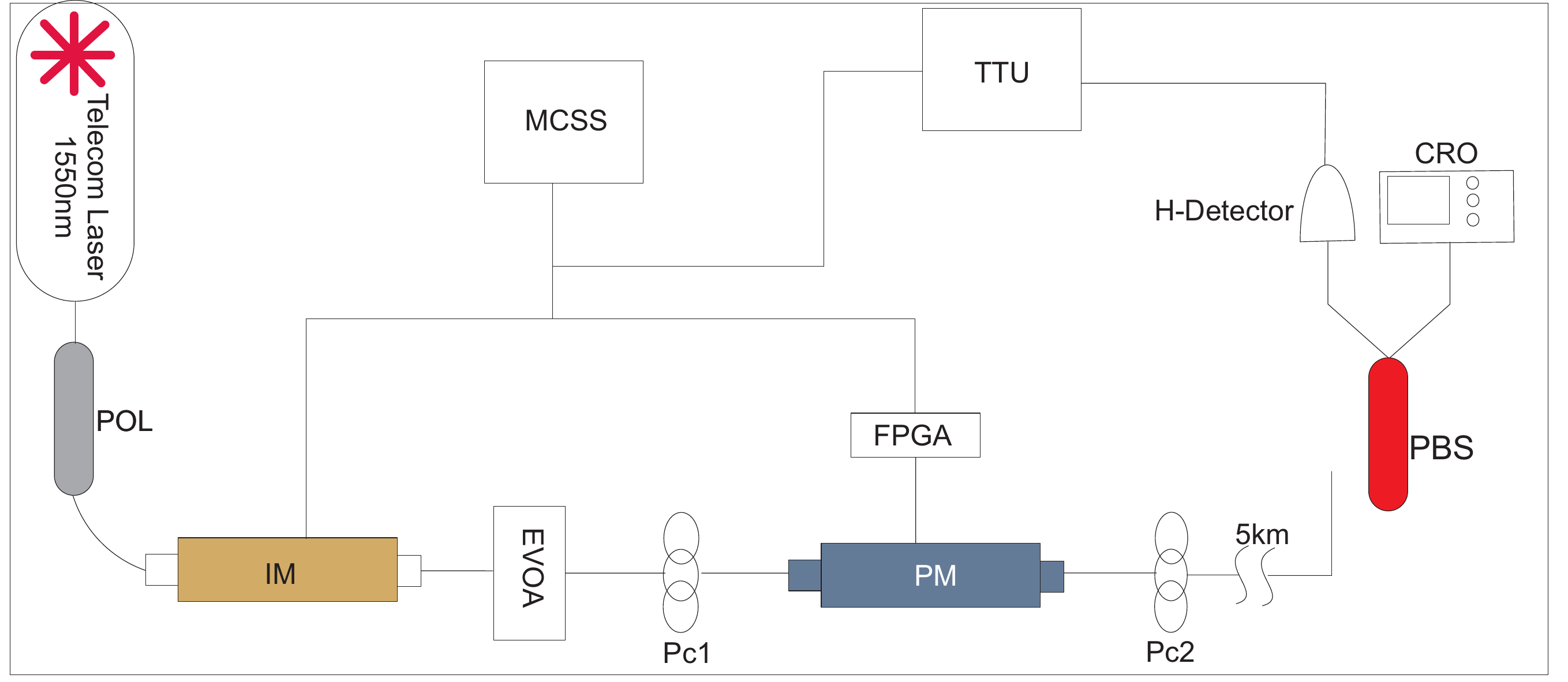}
	\caption{The schematic diagram for the experimental layout is shown. POL is a linear polarizer that is used to polarize the incident unpolarized light coming from the laser, followed by an intensity modulator. The other abbreviated blocked components are as follows; Electronic Variable Optical Attenuator (EVOA), PC1 and PC2, Phase Modulator (PM), Field Programmable Gate Array (FPGA), Master Clock Synchronization Signal (MCSS), Time Tagging Unit (TTU), Horizontal Detector (H-detector),Polarizing Beam Splitter (PBS), Cathode Ray Oscilloscop (CRO). }\label{fig:fig1}
\end{figure}

Assuming a secure quantum channel of $ 5$ km, and for initial experimental training to realize the polarization-encoded QKD systems, we demonstrate the key distribution in one polarization basis through low-speed multiphoton pulses. This paper aims to report fundamental experimental challenges and requirements of essential pre-requisite components while implementing the basic QKD protocol between distant parties. However, the system can subsequently be improved to a realistic level by introducing a single photon source, highly efficient single-photon detectors with a minimal dark count rate, low-loss, and polarization-maintaining optical fibers, by compensating birefringence effects in electro-optic modulators, and using synchronous FPGA modules for rapid switching of polarizations in Pockel cells. The system, whose schematic diagram is shown in Fig. \ref{fig:fig1}, includes a continuous wave $1550$ nm butterfly-type laser module followed by a linear polarizer and a $LiNbO_3$ Intensity modulator used to generate the optical pulses. These pulses are synchronized by the Master Clock Synchronization Signal (MCSS) and are fed to the Phase Modulator (PM) to obtain the desired polarization. After adjusting the Polarization Controller-1 (-2) [PC1 (PC2), the PM recieves two random voltage levels of $0$ and $4$V from the FPGA corresponding to the horizontal and vertical polarization on each cycle of MCSS. In one polarization basis, the detection unit is comprized of a polarizing beam splitter that distinguishes between horizontally and vertically polarized pulses. The horizontal optical pulse (H-pulse) is detected by a Single-Photon Avalanche Detector (SPDAD-Qubitrium InGaAs) while the vertical pulse (V-pulse) is guided towards the Avalanche Photo Detector (APD Thorlabs-PDB415A). The SPAD (H-detector) generates a $4V$ pulse on the registration of each photon and is recorded by a Time Tagging Unit (TTU-model PicoQuant Multiharp $150$), while the result of the APD is visualized in real time on the oscilloscope and it helps to realize the orthogonality condition between H and V pulses as depicted in Fig. \ref{fig:fig3}. The TTU plays a central role and acts as a memory device to record the MCSS, FPGA's random voltages sent to the PM, and the output of the H-detector.

\section{Results and Discussion}\label{sec:TPB}
\begin{figure}[htb]
    \includegraphics[width=\linewidth]{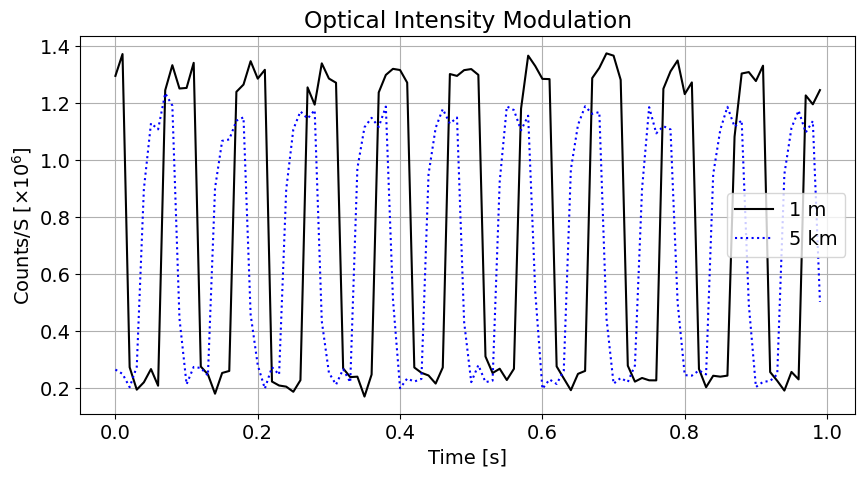}
    \caption{The optical intensity modulation is sketched for two spatial distances, i.e., for $1$ m (black solid curve) and $5$ km (blue dotted curve). The pulses are generated through a $LiNbO_3$ intensity modulator by modulating a Continuous Wave (CW) laser beam. Over a $5$ km fiber link, these pulses experience significantly greater attenuation compared to a $1$ m link, resulting in the subsequent reduction in their peak power.}
    \label{fig:fig2}
\end{figure}

 We have employed a $10$Hz Master Clock Synchronization Signal (MCSS) that synchronizes the intensity modulator, Lithium Niobate (LiNo3) Phase Modulator (PM), and also acts as a reference clock for the FPGA. IM (Model-thorlabs LNA2322) modulates the optical light intensity fed by the CW laser and generates the optical pulses at the frequency of MCSS. These pulses have the peak power of $1mW$ and thus need to be attenuated as a prerequisite. For this, an Electronic Variable Optical Attenuator (EVOA-Thorlabs EVOA1550F) whose attenuation level is adjusted keeping in view the length of the quantum channel (QC). It attenuates the optical power to the level of  -$ 21.55$ dBm for $ 5$ km QC length as shown in Fig \ref{fig:fig2}. Fig. \ref{fig:fig2} explicitly demonstrates that for $5$km long fiber, peak photonic counts are $\approx$ $1.2\times10^6$ per second and are roughly $15\%$ lower compared to the case of $1$m long channel due to fiber losses. The intensity modulation is used to obtain the optical pulses such that each pulse is associated with one of the orthogonal polarization bases. One can set the orthogonality of two polarization bases (say H and V) by manually adjusting the paddles of PC1 and PC2 and through the applied voltages to the PM acting as a pockel cell.
 \begin{figure}[htb]
	\includegraphics[width=\linewidth]{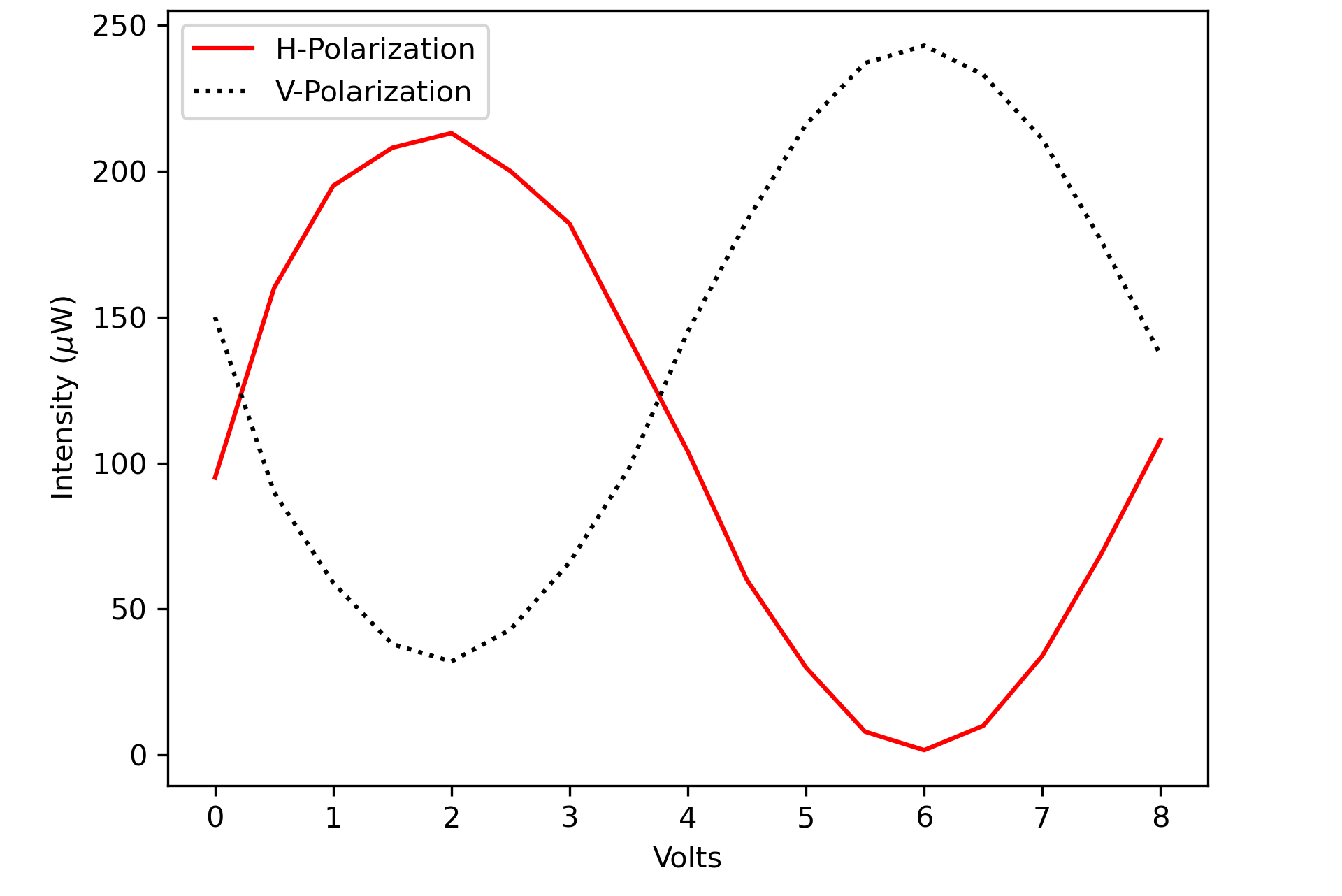}
	\caption{The working of the phase modulator and complementarity of horizontally and vertically polarized light is shown. The intensity of horizontal (vertical) light plotted against applied voltage to the phase modulator is a solid red(dotted block) curve. This figure demonstrates the voltage-dependent modulation of polarization components.   }\label{fig:fig3}
\end{figure}
In Fig. \ref{fig:fig3}, it is shown that within the voltage range of from $(2-6)$V, H and V polarizations switch their positions complementarily. Therefore, $6-2=4$V, is essentially required of Alice for active polarization switching between H and V. However, the careful adjustment of the paddles of PC1 and PC2 is needed to obtain the H polarization (maximum photon counts) at $0$V, i.e., $\approx3.1\times10^4$counts/sec. This enables us to achieve the H polarization and is confirmed by measuring it on the Polarization Analyzer (PA-Model SK$010$PA). The voltages $4$V are then fed to the PM for V polarization and as a result,  all photon population disappears from the horizontal detector, leaving it with only the dark counts. In this way, $0$V and $4$V get safely associated with the H and V polarization basis, respectively, and can be verified through PA. 

\begin{figure}[htb]
	\includegraphics[width=\linewidth]{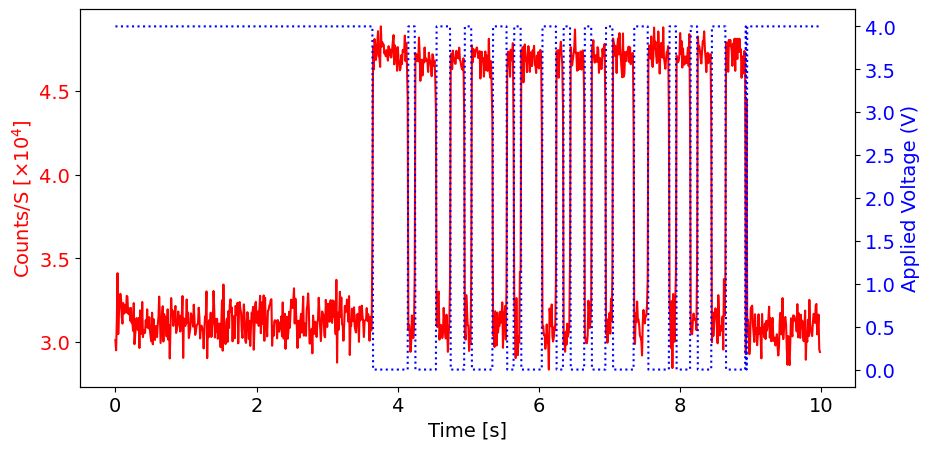}
	\caption{The optical intensity and the applied voltages to the phase modulator on the respective left and right vertical scale are plotted as a function of time. This is purely phase modulation of CW laser beam, and by turning off the intensity modulator.  }\label{fig:fig4}
\end{figure}

 It is important to synchronize the PM with the optical pulses generated from IM in response to MCSS. This requires that the two calibrated voltages ($0$ and $4$ V) should be applied randomly to PM on each cycle of MCSS to achieve the horizontal and vertical polarizations on the sender's end. For this, an FPGA (Spartan $6$ XC$6$SLX$9$) is used that receives the MCSS in parallel with the IM, generates $0$V and $4$V randomly, and records them on each MCSS's cycle. Fig. \ref{fig:fig4} shows this phase modulation behavior and explains how the voltage applied to the PM switches the polarizations between H and V, resulting in two levels of photon counts.  

The PM exploits the phenomenon of birefringence in the lithium niobate crystal that offers different refractive indices to the H and V components of light and induces a phase shift $\phi$ between them. The applied voltage modulates the refractive index and consequently manipulates the $\phi$ in the following way; 
\[
\ket\psi = \cos(\frac{\phi}{2}) \ket{H} + \sin(\frac{\phi}{2}) \ket{V}.
\]\label{Eq1}
The above Equation shows that at $\phi=0$ or $V=0V$ ($\phi=\pi$ or $V=4V$), we get H-polarized (V-polarized) light. In Fig. \ref{fig:fig4}, it is marked that the voltage level $0V$ ($4$V) modulates the refractive index of the $LiNbO_3$ crystal inside the PM, inducing the phase shift $\phi=\pi$ ($\phi=0$) between the H and V component of incoming CW light. Since the detector is coupled only to the H-polarized arm of the Polarizing Beam Splitter (PBS) and therefore, it measures maximum counts i.e., $\approx4.5\times10^4$ counts/sec for $\phi=0$ or $V=0V$ and minimum counts i.e., dark counts for $\phi=\pi$ or $V=4V$ as shown in Fig. \ref{fig:fig4}. It is important to note that this phase modulation is carried out without intensity modulation by keeping the IM in the turned-off position. However, while running the protocol in the next results, IM generates the optical pulses at the frequency of MCSS, and these pulses are ordered strictly synchronized with PM through the FPGA. 
\begin{figure}[htb]
    \centering
    \begin{subfigure}{\linewidth} 
        \includegraphics[width=\linewidth]{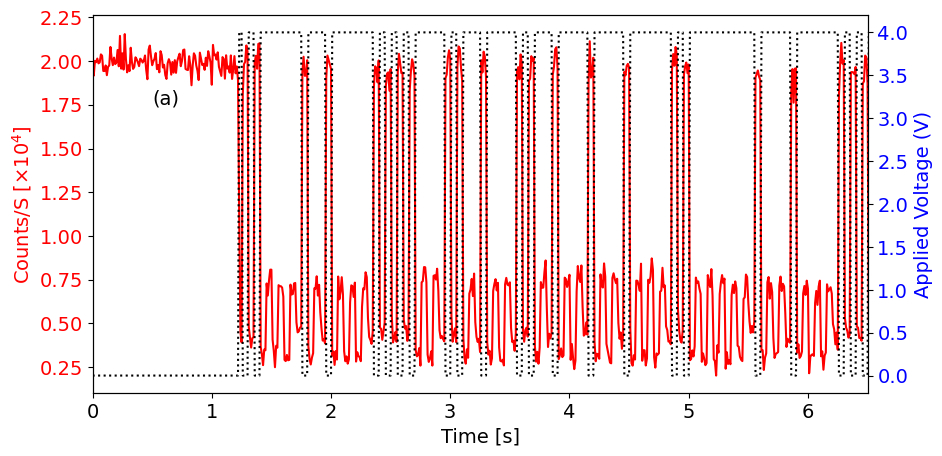}
        \caption{At a transmission rate of 10 Hz.}
        \label{fig:sub1}
    \end{subfigure}

    \vspace{1em}  

    \begin{subfigure}{\linewidth} 
        \includegraphics[width=\linewidth]{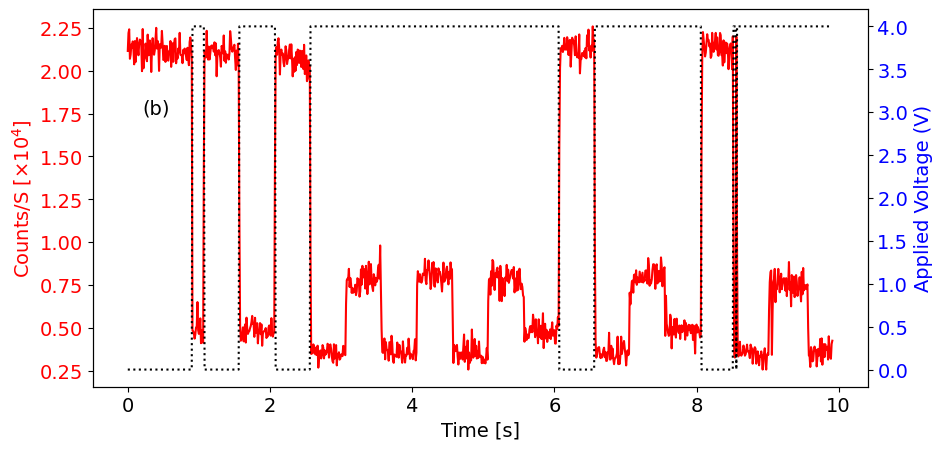}
        \caption{At a transmission rate of 1 Hz.}
        \label{fig:sub2}
    \end{subfigure}

    \caption{The phase modulation of optical pulses, received from the intensity modulator, is illustrated. Alice transmits optical pulses with distinct polarization states—either horizontal or vertical—by randomly applying a voltage from the set [$0$, $4$V] to the phase modulator upon the arrival of each pulse. (a) and (b) illustrate the corresponding photon detection events at Bob’s side.}
    \label{fig:fig5}
\end{figure}

In Fig. \ref{fig:fig5}, the ultimate detection at Bob's end is visualized at different frequencies of $10$Hz and $1$Hz in \ref{fig:fig5} (a) and  \ref{fig:fig5} (b), respectively. It demonstrates that by switching on the MCSS after about $\approx1.2$ seconds, IM and PM start modulating the optical intensity. However, the modulation of PM is random as it depends upon the FPGA's output ($0V$ or $4V$ randomly) on each cycle i.e., the detector measures the $\approx2\times10^4$ counts on $0V$ (H polarization) and $\approx7.5\times10^3$ counts for the instances of $4V$ (V Polarization). Since MCSS is a unipolar square wave signal of $4V$ amplitude, therefore, IM cuts off the light at peak voltage ($4V$), and the detector measures only dark counts i.e., $\approx2.5\times10^3$ at that instance as shown in Fig.\ref{fig:fig5}.

At Alice's end, We generate a binary key from an indigenously developed Quantum Random Number Generator (QRNG) \cite{haider2023quantum} and load it into the FPGA. As mentioned earlier, FPGA is programmed in such a way that $0$s and $1$s of the binary key are converted into $0V$ and $4V$s, respectively. These two voltage levels are used to distinguish between H and V polarizations states. The orthogonality of H and V states is set maximum limit through the PC1 and PC2 and by making sure that the ratio of horizontal to the dark counts of the H-detector should be maximum. In other words, fidelity of H and V polarization state is kept at maximum by using both polarization controllers. At Bob's end, detection data from TTU accompanied by the data of MCSS is analyzed and two levels of counts above the dark counts from the data are duly separated by associating them with the H or V polarization ($0$ or $1$ of the binary key, respectively).
\begin{figure}[htb]
	\includegraphics[width=\linewidth]{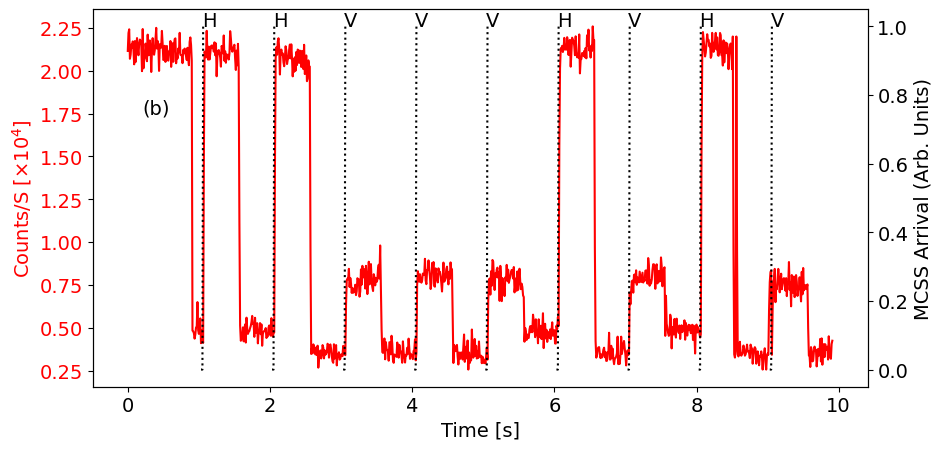}
	\caption{Illustration of how Bob extracts his key information from the received data. As he is using only an H-detector, the pulse whose polarization is vertical has much lower counts than its horizontal counterpart. This leaves the clear and distinguishable signature of both polarizations and can be recorded for further post-processing. }\label{fig:fig6}
\end{figure}

The temporal resolution of TTU is kept $1\times10^{-2}$s that takes $100$ data counts samples after each second. It is essential to avoid repetition in the counts and therefore, the MCSS is also directly measured by another channel of TTU as a reference. Further, only those data samples are chosen for which MCSS arrives as shown in Fig. \ref{fig:fig6}. This strategy shrinks the large data samples into one sample per cycle of the MCSS and helps to generate the raw key. The nine samples, for example, are retained based on the arrival of MCSS and the count rate level of $\approx 2\times10^4$ ($\approx 7.5\times10^3$) can be linked to the H (V) polarizations measured by the H-detector in Fig. \ref{fig:fig6}. Ideally, the H-detector must be at the dark count level for V-polarization instances, however, it is the fidelity error between H and V states that induces the non zero count rate in addition to the above-mentioned dark counts. This fidelity error can be minimized by fine adjustment of PC1 and PC2, despite the fact that it cannot be completely removed due to the uncontrollable polarization drift in fibers.
\begin{figure}[htb]
	\includegraphics[width=\linewidth]{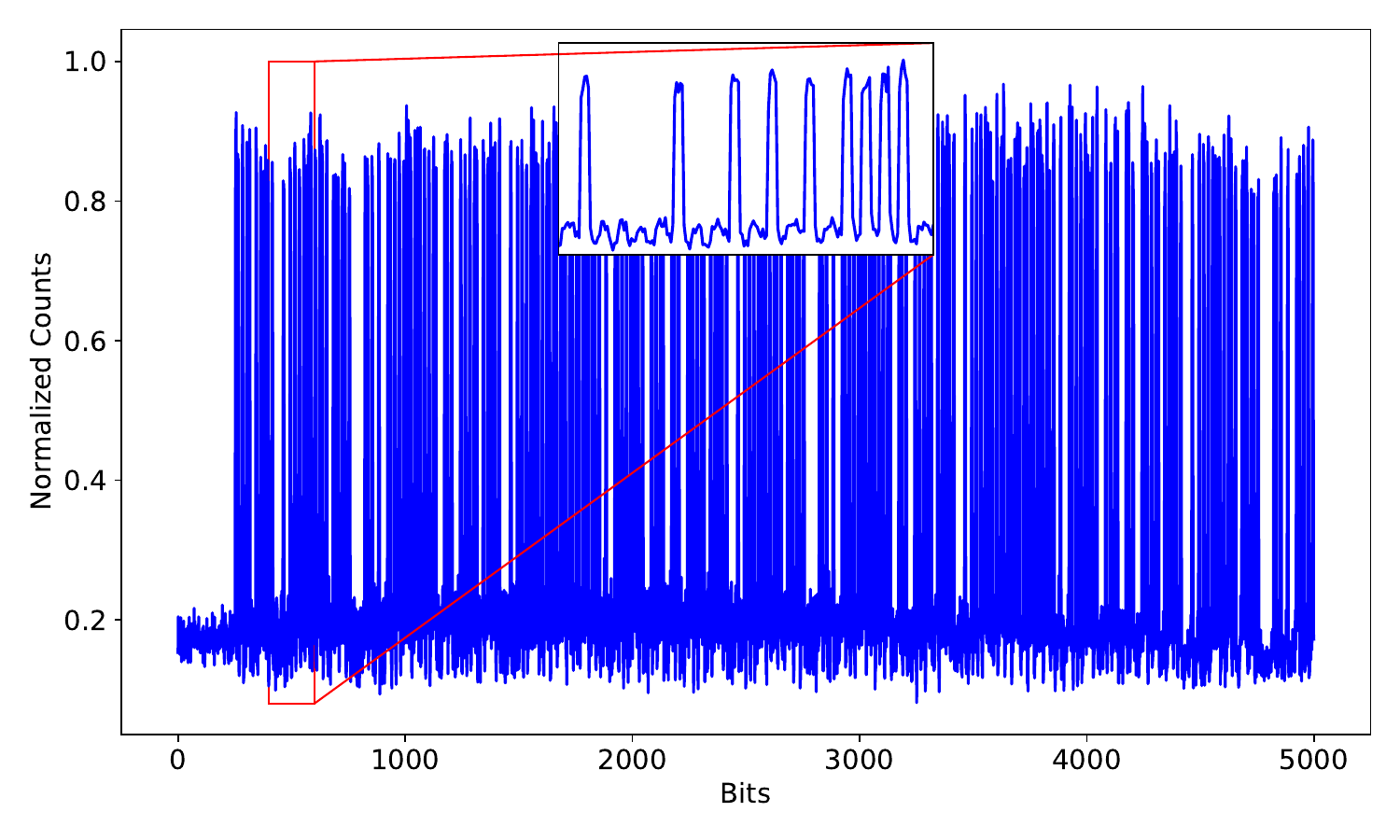}
	\caption{The real-time detection result at the receiver end for the first $ 5000$ bits/pulses is shown. Each pulse is counted as a bit, and normalized photon counts are plotted against the bit index, with reference to the MCSS. The zoomed portion of the plot (inset) displays the randomness of the horizontally polarized light pulses, while vertically polarized pulses are at the dark count level of the detector. }\label{fig:fig7}
\end{figure}
The known length of the binary key is then loaded to the FPGA and triggered by MCSS to start sending to the PM. At the same time, H-detector starts recording the counts and is visualized in Fig. \ref{fig:fig7}. It is shown in Fig. \ref{fig:fig7} (inset) that horizontal pulses are detected randomly and there is no H-pulse overlap at the instances of vertical polarization. This enables one to extract the data reference to the MCSS and to make the binary raw key at Bob's end. The frequency of MCSS i.e., $10$Hz defines the key rate which can be systematically enhanced for high key rate applications. As we are working in one polarization basis (H and V) so it eleminate the step of sifting in our model as bob is using true polarization basis every time. Finally a part of the key is chunked and shared between them to estimate the QBER as shown in Fig. \ref{fig:fig8}. The system is operated for $5$ minutes and $3000$ bits are shared between Alice and Bob on $5$km long fiber link. It is worth noting that no bit is fliped for $2.5$ minutes and QBER remained at zero. Later on, polarization drift reduces the contrast between horizontal and vertical pulses and induces an error as shown in Fig. \ref {fig:fig7}.
\begin{figure}[htb]
	\includegraphics[width=\linewidth]{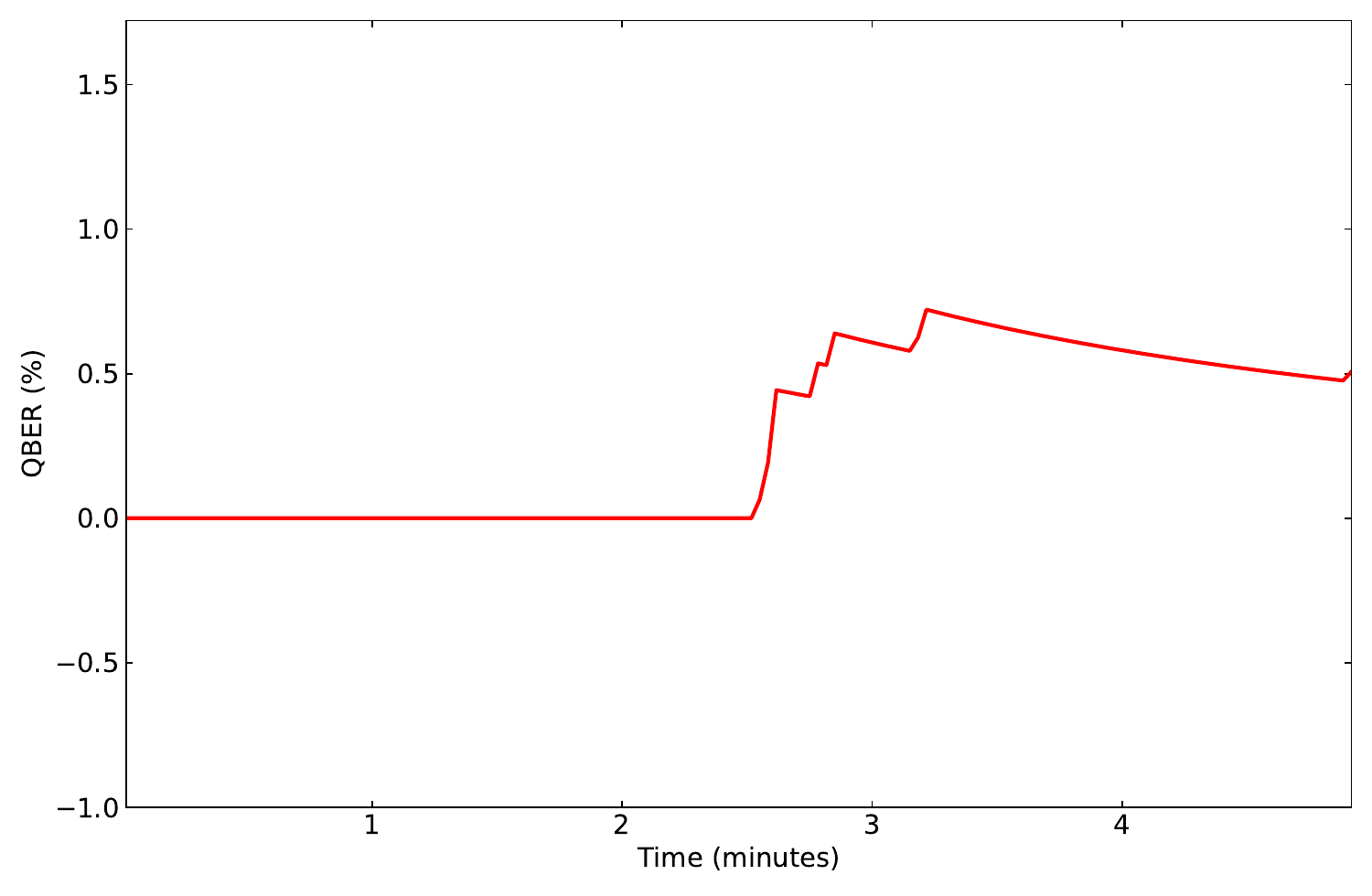}
	\caption{QBER plotted as a function of time (in minutes). During the first $2.5$ minutes, the contrast between horizontally and vertically polarized light is maximal, resulting in zero bit-flip errors; hence, the QBER remains at zero. Thereafter, the QBER gradually increases but remains below $1$ \% throughout the observation period. }\label{fig:fig8}
\end{figure}
\section{Conclusion}\label{sec:conc}
We have experimentally demonstrated the fundamental setup for polarization encoding at laboratory scale and its associated requirements and challenges that one needs to cater while starting a realistic experiment. For a $ 5$ km long fiber link, we have used the optical pulses at the frequency of $10$Hz that carry the polarization information from the sender to the receiver. Each pulse is designated as a binary bit by modulating it with one of the two orthogonal polarization basis. For the first $2.5$ minutes of the experiment, all bits are successfully and distinguishablly distributed. The birefringence in the crystals and optical fibers, the temperature destabilization, and other drifts in fibers reduces the orthogonality index between the polarizations and cause the error rate in the next $2.5$ minutes. However, this error is small enough ($<1\%$) and can be corrected by employing the basis error correction techniques \cite{dixon2014high,gumucs2021novel}. We believe that this basic and primary guide for the implementation of polarization encoding will contribute in simplifying the setups related to secure communication, including QKD systems.

\bibliography{ref.bib}
\end{document}